\documentclass[11pt,a4paper]{article}
\usepackage[marginratio=1:1,height=600pt,width=460pt,tmargin=100pt]{geometry}
\usepackage[utf8]{inputenc}
\usepackage[parfill]{parskip}    % Activate to begin paragraphs with an empty line rather than an indent
\usepackage{url}            % simple URL typesetting
\usepackage{booktabs}       % professional-quality tables
\usepackage{amsfonts}       % blackboard math symbols
\usepackage{nicefrac}       % compact symbols for 1/2, etc.
\usepackage{microtype}      % microtypography
\usepackage[dvipsnames]{xcolor}
\usepackage{graphicx}
\usepackage{amsmath}
\usepackage{amssymb}
\graphicspath{{./pics/}}
\usepackage{subcaption}
\usepackage{amsthm}
\usepackage{amscd}
\usepackage{mathrsfs}
\usepackage{bm}
\usepackage{algorithmic, algorithm}
\usepackage{multirow}
\usepackage{mathtools}
\usepackage{hyperref}
\usepackage{url}
\usepackage{enumitem}
\usepackage{authblk}
\usepackage{overpic}
\usepackage[normalem]{ulem}

\allowdisplaybreaks

\newcommand{\R}{\mathbb{R}}
\newcommand{\N}{\mathcal{N}}
\newcommand{\T}{\mathcal{T}}
\newcommand{\CL}{\mathcal{L}}
\newcommand{\bx}{\bm{x}}

\newcommand{\bu}{\bm{u}}
\newcommand{\btheta}{\bm{\theta}}

\newtheorem*{thm*}{Theorem}

\newtheorem*{lemma*}{Lemma}

\usepackage{array}
\newcolumntype{C}[1]{>{\centering\let\newline\\\arraybackslash\hspace{0pt}}m{#1}}

\title{Discovering Governing Equations in Discrete Systems Using PINNs}

\author[1]{Sheikh Saqlain}
\author[1]{Wei Zhu}
\author[2]{Efstathios G.~Charalampidis}
\author[1]{Panayotis G.~Kevrekidis}
\affil[1]{Department of Mathematics and
Statistics, University of Massachusetts
Amherst, Amherst, MA 01003-4515, USA}
\affil[2]{Mathematics Department,
California Polytechnic State University, 
San Luis Obispo, CA 93407-0403, USA}

\date{}

\begin{document}
\maketitle

\begin{abstract}
Sparse identification of nonlinear dynamical systems
is a topic of continuously increasing significance in
the dynamical systems community. Here we explore it
at the level of lattice nonlinear dynamical systems
of many degrees of freedom.~We illustrate the ability
of a suitable adaptation of Physics-Informed Neural
Networks (PINNs) to solve the inverse problem of
parameter identification in such discrete, high-dimensional
systems inspired by physical applications. The 
methodology is illustrated in a diverse array
of examples including real-field ones ($\phi^4$ and sine-Gordon), as well as complex-field (discrete
nonlinear Schr{\"o}dinger equation) and going
beyond Hamiltonian to dissipative cases (the 
discrete complex Ginzburg-Landau equation). 
Both the successes, as well as some limitations of
the method are discussed along the way.
\end{abstract}

\section{Introduction}
% ======================================================================

While the study of nonlinear partial differential equations (PDEs) captures
the lion's share of modeling efforts in physical, chemical and biological
problems, the study of nonlinear dynamical lattices has received growing
attention over the last decades~\cite{Aubry06,Flach:2008,FPUreview,pgk:2011}.
To some degree, this interest stems from the consideration of discretization
methods for simulating PDEs. However, arguably, the most appealing aspect
of such lattice problems is that they naturally emerge as the 
suitable model in systems where there is a degree of ``granularity''/lattice
structure. This may stem from waveguides and their arrays in 
nonlinear optics~\cite{moti}, or  in experiments of micromechanical oscillator arrays~\cite{sievers} %
as well as lattice nonlinear electrical circuits~\cite{remoissenet}. It may arise in material
science systems~\cite{yuli_book,granularBook}, in antiferromagnetic~\cite{lars3},
or more generally anharmonic~\cite{ST,page} crystals, in superconducting
settings of Josephson-junction ladders~\cite{alex,alex2} or in biological
models of DNA base pairs~\cite{Peybi,Yomosa1983}. Such lattice models may also be
effective ones, emulating the periodic variation of optical lattices
in atomic condensates~\cite{Morsch}. 

On the other hand, over the past few years there has been an explosion
of interest in the use of data-driven techniques towards the study of
physical phenomena and the development, as well as identification of
relevant models~\cite{karniadakis2021physics}. Among the most dominant
methodologies in that regard for the solution of both inverse and forward
problems in PDEs have been methodologies
such as PINNs (Physics-Informed Neural Networks)~\cite{raissi_physics-informed_2019}, and the subsequent
extension of DeepXDE~\cite{lu2021deepxde}, as well as the SINDY
(sparse identification of nonlinear systems) method of~\cite{brunton_discovering_2016}, sparse optimization in \cite{schaeffer2017learning}, meta-learning \cite{feliu2020meta}, and neural operators \cite{
li2021fourier}. There have been numerous variations
and extensions of these approaches in a wide range of problems
(a small subset of which, e.g., 
contain~\cite{sirignano2018dgm,weinan2018deep,gu2021selectnet,shin2020error,luo2020two}). Yet, it can be argued that these approaches have been, by and large,
limited to continuum PDE problems and the emergent aspect of nonlinear 
dynamical lattices has been somewhat overlooked. 

Here, we build on the earlier work of some of the present authors~\cite{zhu_neural_2022}, which 
was aiming to build in the neural networks more
of the physical structure of the underlying problem (in that case
through symmetries, while other authors have also enforced, e.g., the 
symplectic structure of potential underlying systems~\cite{george_again}). 
Our emphasis in the present work is to adapt methods of the above type, most notably
PINNs, to nonlinear dynamical lattices. In particular, we will select a sequence
of progressively more complex yet physically relevant examples and seek to leverage
the above computational methodology, albeit now in an inherently discrete, high-dimensional
setting.  
By high-dimensional here, we refer to the number of 
degrees of freedom (and not the spatial dimension of
the problem).
Our aim will be  to solve the inverse problem of the identification of linear and nonlinear
coefficients of the models building progressively from simpler to more complex. 
We will start from a real $\phi^4$ discrete nonlinear Klein-Gordon 
system~\cite{p4book} and subsequently move to the complex variant of
the model, namely the discrete nonlinear Schr{\"o}dinger (DNLS) model~\cite{kevrekid_dnls_book}. We will subsequently explore an
example bearing a different type of complexity where the nonlinearity
is not a pure power law, but rather a sinusoidal one in the form of the
Frenkel-Kontorova~\cite{braun1998} or discrete sine-Gordon~\cite{SGbook}
nonlinearity. This will serve to showcase some of the challenges and
limitations of the approach. Finally, we will extend considerations
beyond the Hamiltonian class of examples to the discrete variant~\cite{PhysRevE.67.026606,GL2}  of
the Ginzburg-Landau equation~\cite{RevModPhys.74.99}, a topic that continues
to be of wide interest in its own right as evidenced in the recent
review of~\cite{Salerno2022}.

Our presentation will be structured as follows. In Section 2, we
will provide some of the mathematical background of the problem, both
at the level of the dynamical models under consideration (Sec.~\ref{sec:discrete_systems}) and as concerns 
the PINN approach  (Sec.~\ref{sec:pinns}).
Then, upon explaining how to adapt the discovery
of the governing equation to nonlinear dynamical
lattices in Sec.~\ref{sec:disc_pinns}, we present
our numerical experiments in Sec.~\ref{sec:num_exp}.
Finally, in Sec.~\ref{sec:concl}, we summarize
our findings and present a number of possibilities
for future studies.

\section{Background}

\subsection{Discrete nonlinear lattices}
% ======================================================================
\label{sec:discrete_systems}

In this work, we consider a variety of 1D discrete nonlinear lattices 
consisting of $N$ nodes. In all the cases that we will focus on hereafter, 
$u_{n}(t)$ (which can be real or complex, depending on the model) will 
correspond to a dynamical variable with $n=1,\dots,N$. We start our presentation 
of the models by considering first 
the discrete $\phi^{4}$ model~\cite{p4book}
\begin{align}
\ddot{u}_{n} = C(u_{n+1} + u_{n-1} - 2u_n) + 2(u_n - u_n^3), \quad u_{n}\in\mathbb{R},
\label{dphi4}
\end{align}
where the overdot stands for the temporal derivative of $u_{n}$, and $C=1/h^{2}(>0)$ effectively represents
the coupling constant with $h$ representing the lattice spacing between adjacent nodes. 
It should be further noted that neighboring sites in Eq.~\eqref{dphi4} are coupled due 
to the presence of the $(u_{n+1}+u_{n-1}-2u_n)$ 
discrete Laplacian term therein, and the strength of the 
coupling is dictated by the magnitude of $C$. That is, a large value of $C$, i.e., $C\gg 1$
or equivalently $h\ll 1$ signifies that Eq.~\eqref{dphi4} 
is close to the continuum $\phi^4$
limit, whereas a small value of $C$  will result in a highly discrete
system. Moreover, this coupling term (involving $C$), that will be ubiquitous in all of 
the models that we consider herein, emanates from the discretization of the Laplacian operator 
in 1D by using a centered, second-order accurate finite difference scheme. This provides a vein along which
the model can interpolate between the so-called 
anti-continuum limit~\cite{Aubry06} of $C=0$ and
the continuum limit of the respective PDE.
Eq.~\ref{dphi4} is a model, variants of which have been
useful towards understanding solitary wave
dynamics in a simpler, real lattice nonlinear
system~\cite{PhysRevE.76.026601,PhysRevE.72.035602}.
We thus use it as a preamble to studying the
complex DNLS variant of the model. 

Indeed, we subsequently focus on
the well-known yet physically relevant, discrete 
nonlinear Schr\"odinger (DNLS) equation~\cite{kevrekid_dnls_book} with a focusing 
(cubic) nonlinearity:
\begin{align}
i\dot u_n = -C(u_{n+1}+u_{n-1}-2u_n) -|u_n|^2u_n, \quad u_{n}\in\mathbb{C}.
\label{dnls}
\end{align}
Here, we allow the relevant field representing, e.g.,
the envelope of the electric field along an optical
waveguide array~\cite{moti} or the quantum-mechanical
wavefunction along the nodes of a deep optical
lattice~\cite{Morsch}, to be complex.

Another intriguing example consists of the discrete sine-Gordon (DsG)~\cite{SGbook}, also known
in dislocation theory as Frenkel-Kontorova 
model~\cite{braun1998}:
\begin{align}
\ddot{u}_{n} = C(u_{n+1} + u_{n-1} - 2u_n) -\sin{(u_n)}, \quad u_{n}\in\mathbb{R}.
\label{dsG}
\end{align}
This model, similarly to Eq.~(\ref{dphi4}) admits
kink-like solutions. However, it also has a key
distinguishing feature from the former. Namely,
it bears a transcendental nonlinear function, 
one that cannot be expressed as a simple power
law. Indeed, the difficulty to represent
such a simple pendulum (unless further, e.g.,
Hamiltonian structure
of the problem is built in the sparse identification
approach) has been previously documented, e.g.,
as concerns SINDY in~\cite{pmlr-v190-lee22a}.
%
%
%both admitting ``kink''-like solutions~\cite{kevrekid_jesus_book}.

Finally, the other fundamental model of interest in the present work is the discrete, complex 
Ginzburg-Landau (DCGL) equation:
\begin{align}
\dot u_n = (1+i)C(u_{n+1} + u_{n-1} - 2u_n) - (1-i) |u_n|^2u_n + u_n, \quad u_{n}\in\mathbb{C},
\label{dcgl}
\end{align}
with a cubic nonlinearity~\cite{PhysRevE.67.026606}. The DCGL can be considered
as a (dissipative) perturbation of the DNLS [cf. Eq.~\eqref{dnls}]. Such settings are of interest in the same
contexts as DNLS when dissipative perturbations
are present, as, e.g., in experimental studies such as that of~\cite{doi:10.1126/science.abf6873} in optics or the one of~\cite{doi:10.1126/sciadv.aat6539} in atomic
Bose-Einstein condensates.
%many contexts ranging from coupled waveguides and lasers to Bose-Einstein condensates 

%(BECs). 
%Again, the coupling term in Eq.~\eqref{dcgl} corresponds to the discrete Laplacian 
%(with $C=1/h^{2}$ as before). Alongside the DNLS and DCGL %models, we will consider 

\begin{table}[pt!]
\centering
\resizebox{\columnwidth}{!}{%
\begin{tabular}{|c|c|c|c|}
\hline
Model             & Equation      & IC($u$) \\ \hline
DNLS              &   $\displaystyle{i\dot u_n = -C(u_{n+1}+u_{n-1}-2u_n) -|u_n|^2u_n}$  & $\displaystyle{e^{-x_{n}^2}}$ \\ \hline
DCGL              &   $\displaystyle{\dot{u}_n = (1+i)C(u_{n+1} + u_{n-1} - 2u_n) - (1-i) |u_n|^2u_n + u_n}$  & $\displaystyle{\tanh{\!x_{n}}\exp{(i\ln{(\cosh{x_{n}}}))}}$ \\ \hline
Discrete $\phi^4$ & $\ddot{u}_{n} = C(u_{n+1} + u_{n-1} - 2u_n) + 2(u_n - u_n^3)$ 
    &  $\tanh{\left(\frac{x_{n}}{\sqrt{1-v^2}}\right)}$\\ \hline
DsG               &   $\displaystyle{\ddot{u}_{n} = C(u_{n+1} + u_{n-1} - 2u_n) -\sin{(u_n)}}$ 
                  &  $\displaystyle{4\arctan{\left(\exp{\left(\frac{x_{n}}{\sqrt{1-v^2}}\right)}\right)}}$ \\ \hline    
\end{tabular}
}
\caption{Discrete nonlinear lattices that are considered in this work together
with the respective initial conditions. Note that $x_{n}$ are grid
points with $n=1,\dots,N$, taken uniformly from the interval 
$\left[-\frac{N}{2\sqrt{C}},\frac{N}{2\sqrt{C}}\right]$.
}
\label{our_models}
\end{table}

A recap of the principal models of interest
can be found in the Table~\ref{our_models}.
The table contains not only the mathematical form
of each of the models but also the initial conditions (ICs)
that are used therein in order to perform the model
training (cf.~Section~\ref{sec:data_generation}).
We conclude this section by mentioning the boundary conditions (BCs) that we will employ
for all the above models. In particular, we impose free BCs at both ends of the lattice, 
i.e., $u_{0}=u_{1}$ and $u_{N+1}=u_{N}$. These BCs can be thought of as the discrete 
analogues of zero Neumann BCs in the continuum limit, and emanate through the discretization 
of the latter through first-order accurate, forward and backward finite difference formulas, 
respectively. Having discussed about the models of interest herein, we now turn into a brief 
overview of Physics-Informed Neural Networks (PINNs).

\subsection{Physics-Informed Neural Networks}
% ======================================================================
\label{sec:pinns}
Since their introduction by Raissi et al. \cite{raissi_physics-informed_2019}, PINNs have garnered growing attention from the scientific machine learning community 
due to their flexible and gridless design in data-driven modeling of forward and inverse problems. 
Consider, for instance, the following parametrized PDE:
\begin{align}
  \label{eq:general-pde}
  \left\{
  \begin{aligned}
    & u_t = \mathcal{N}(u; \lambda), \quad &&  \bx\in\Omega, t\in [0, T],\\
    & u(\bx, 0) = g(\bx), \quad &&  \bx\in\Omega,\\    
    & \mathcal{B}u(\bx, t) = h(\bx,t), \quad &&  \bx\in\partial\Omega, t\in [0, T],
  \end{aligned}\right.
\end{align}
where $u(\bx, t)$ is the unknown, $\mathcal{N}(\cdot;\lambda)$ is a (spatial) nonlinear 
differential operator parametrized by $\lambda$, and $\mathcal{B}$ is an operator associated 
with a specific BC.

In \textit{forward problems}, i.e., when the model 
parameter $\lambda$ is fixed and given, one aims to derive the (numerical) solution $u(\bx, t)$ 
of Eq.~\eqref{eq:general-pde} with the specified initial and boundary conditions. A PINN for 
Eq.~\eqref{eq:general-pde} in this setting is a neural network ansatz $\hat{u}(\bx, t;\btheta)$ that serves 
as a surrogate of the solution $u(\bx, t)$, where $\btheta$ is the collection of all trainable 
parameters of the neural network, e.g., weights and biases of a fully-connected feed-forward PINN. The optimal solution $\hat{u}(\bx, t;\btheta^*)$ 
is searched such that the constraints imposed by the PDE and the initial/boundary conditions are 
(approximately) satisfied. More specifically, let $\T_{\N}\subset \Omega\times [0, T]$, $\T_{g}\subset \Omega\times \{t=0\}$ 
and $\T_h\subset \partial\Omega\times[0, T]$ be three finite collections of scattered ``training'' 
points sampled from their corresponding regions.~The discrepancy between $\hat{u}(\bx, t;\btheta)$ 
and the constraints in Eq.~\eqref{eq:general-pde} is measured through the following loss function 
$\CL(\btheta; \T_\N, \T_g, \T_h)$ defined as a weighted sum of the discrete $l^2$ norms of the 
residuals for the PDE and the initial/boundary conditions:
\begin{align}
  \label{eq:pinn-res-min-forward}
  \CL(\btheta; \T_\N, \T_g, \T_h) \coloneqq w_\N \CL_\N(\btheta; \T_\N) + w_g \CL_g(\btheta; \T_g) + w_h \CL_h(\btheta; \T_h),
\end{align}
where
\begin{align}
  & \CL_\N(\btheta; \T_\N) = \frac{1}{|\mathcal{T}_\N|}\sum_{(\bx, t)\in\T_\N}\left|\hat{u}_t(\bx, t;\btheta)-\N(\hat{u};\lambda)(\bx, t;\btheta) \right|^2,\\
  & \CL_g(\btheta; \T_g) = \frac{1}{|\T_g|}\sum_{(\bx, 0)\in \T_g}\left|\hat{u}(\bx, 0;\btheta)-g(\bx)\right|^2,\\
  & \CL_h(\btheta; \T_h) = \frac{1}{|\T_h|}\sum_{(\bx, t)\in \T_h}\left|\mathcal{B}\hat{u}(\bx, t;\btheta)-h(\bx,t )\right|^2,
\end{align}
$|\T_\N|,|\T_g|,|\T_h|$ are the cardinalities of the sets $\T_\N, \T_g, \T_h$, and $w_\N, w_g, w_h > 0$ are the weights. The differential operators in the loss function $\CL(\btheta; \T_\N, \T_g, \T_h)$ 
are obtained through automatic differentiation \cite{baydin2018automatic}, and  
$\btheta^* = \arg\min_{\btheta} \CL(\btheta; \T_\N, \T_g, \T_h)$ is typically solved by gradient-based optimization 
methods (such as ADAM \cite{DBLP:journals/corr/KingmaB14} or L-BFGS \cite{liu1989limited}).

In \textit{inverse problems}, the parameter $\lambda$ in Eq.~\eqref{eq:general-pde} is not known, and 
the objective is to infer the unknown parameter $\lambda$ from some extra measurement of the system in 
addition to the PDE and the initial/boundary conditions. For instance, let $\T_{f}\subset \Omega\times [0, T]$, 
and assume that the values of the solution $u(\bx, t)$ are known for $(\bx, t)\in\T_f$:
\begin{align}
  \label{eq:inverse_observation}
  u(\bx, t) = f(\bx, t), \quad \forall (x, t)\in\T_f\subset  \Omega\times [0, T].
\end{align}
In this setting, the loss function $\CL(\btheta, \lambda; \T_\N, \T_g, \T_h, \T_f)$ will have an extra 
term corresponding to the additional information of the system given by Eq.~\eqref{eq:inverse_observation}:
\begin{align}
  \label{eq:pinn-res-min-inverse}
    \CL(\btheta, \lambda; \T_\N, \T_g, \T_h, \T_f) \coloneqq w_\N \CL_\N(\btheta, \lambda; \T_\N) %
    + w_g \CL_g(\btheta, \lambda; \T_g) + w_h \CL_h(\btheta, \lambda; \T_h) + w_f  \CL_f(\btheta, \lambda; \T_f),
\end{align}
where
\begin{align}
  & \CL_f(\btheta, \lambda; \T_f) = \frac{1}{|\mathcal{T}_f|}\sum_{(\bx, t)\in\T_f}\left|\hat{u}(\bx, t;\btheta)-f(\bx, t) \right|^2.
\end{align}
Another notable change from Eq.~\eqref{eq:pinn-res-min-forward} to Eq.~\eqref{eq:pinn-res-min-inverse} 
is that the unknown $\lambda$ also becomes the trainable parameter of the model, and it is jointly searched with $\btheta$ 
by minimizing Eq.~\eqref{eq:pinn-res-min-inverse}.

\section{Discovering governing equations in discrete systems}
\label{sec:disc_pinns}
% ======================================================================
The problem of interest to us herein concerns the data-driven discovery of governing 
equations, and in particular, the ones corresponding to the nonlinear dynamical 
lattices discussed in Section~\ref{sec:discrete_systems}. To do so, consider 
a 1D lattice of $N$ nodes with a dynamical variable $u_n(t)\in\mathbb{R}$ (or $\mathbb{C}$) associated to each node $n=1, 2, \dots, N$. Assume that the 
evolution of $\bu = (u_1, u_2, \dots, u_N)$ is governed by the following nonlinear 
dynamics
\begin{align}
\label{eq:general-ode}
    \dot{\bu} = (\dot{u}_1, \dot{u}_2, \dots, \dot{u}_N) = \N(u_1, \dots, u_N),
\end{align}
where $\N:\R^N\to\R^N$ (or $\mathbb{C}^N\to\mathbb{C}^N$) is an operator comprised of 
inter-site couplings between nearest neighbors, but the explicit form of $\N(u_1, \dots, u_N)$ 
is unknown. Our objective is to learn the governing equation $\N$ from sparse (temporal) 
observations of the nonlinear dynamics $\bu(t)=\bm{f}(t)$ at times $t\in \T_{\bm{f}}\subset [0, T]$, where $T>0$ is the terminal time of the system.

The differences between our setting and the PDE inverse problem explained in Section~\ref{sec:pinns} are two-fold. 
First, even though the inter-site couplings in $\N$ between nearest neighbors can sometimes be viewed as finite 
differences, bearing resemblance to their continuous counterparts of (spatial) differential operators (cf.~Section~\ref{sec:discrete_systems}), 
the system described by Eq.~\eqref{eq:general-ode} is intrinsically discrete. Second, unlike the parametric nonlinear 
operator $\N(u;\lambda)$ in Eq.~\eqref{eq:general-pde}, whose explicit dependence on $\lambda$ is given, the governing 
equation [cf. Eq.~\eqref{eq:general-ode}] is generally unknown, except for the prior knowledge that the right-hand-side 
$\N(u_1, \dots, u_N)$ involves only a 
shift-invariant coupling to nearby sites.

We thus make the following modifications to the PINN model of Eq.~\eqref{eq:pinn-res-min-inverse}. For systems with 
real dynamical variables $\bu(t)\in \R^N$, the PINN $\hat{\bu}: \R\to\R^N$ takes only time $t$ as the input, 
which is mapped through an $L$-layer fully-connected neural network to the output corresponding to an $N$-dimensional vector 
$\hat{\bu}(t) = (\hat{u}_1(t),\hat{u}_2(t),\dots, \hat{u}_N(t))\in\R^N$. Since the form of $\N(u_1, \dots, u_N)$ is 
not known, we build an overcomplete library $\mathrm{Lib}= \{D_\alpha\}_{\alpha\in A}$ of shift-invariant 
discrete spatial operators modeling the 
linear 
inter-site couplings between nearest neighbors, as well
as different types of nonlinear contributions. For instance, one dictionary 
element that is included in many of our numerical experiments is the discrete Laplacian
\begin{align}
    (D_2 \bu)_n = u_{n-1} - 2u_n + u_{n+1}.
\end{align}
The unknown operator $\N:\R^N\to\R^N$ is then modeled as a linear combination $\N = \sum_{\alpha\in A}\lambda_\alpha D_\alpha$ 
of elements $D_\alpha$ in the library, and the expansion coefficients $\bm{\lambda} = (\lambda_\alpha)_{\alpha\in A}$ 
are learned by minimizing the loss function
\begin{align}
    \CL(\btheta, \bm{\lambda}; \T_\N, \T_{\bm{f}}) \coloneqq w_\N\CL_\N(\btheta, \bm{\lambda};\T_\N) + w_{\bm{f}}\CL_{\bm{f}}(\btheta, \bm{\lambda};\T_{\bm{f}}),
\end{align}
where
\begin{align}
  & \CL_\N(\btheta, \bm{\lambda}; \T_\N) = \frac{1}{|\mathcal{T}_\N|}\sum_{t\in\T_\N}\left|\dot{\hat{\bu}}(t;\btheta)%
  -\sum_{\alpha\in A}\lambda_\alpha D_\alpha \hat{\bu}(t; \btheta) \right|^2,\\
  & \CL_{\bm{f}}(\btheta, \bm{\lambda}; \T_{\bm{f}}) = \frac{1}{|\mathcal{T}_{\bm{f}}|}\sum_{t\in\T_{\bm{f}}}\left|\hat{\bu}(t;\btheta)%
  -\bm{f}(t; \btheta) \right|^2,
\end{align}
and $\T_\N$, $\T_{\bm{f}}$, respectively, are subsets of $[0,T]$ corresponding to the training 
collocation points at which the ODE residual and the discrepancy between $\hat{\bu}$ and the observed 
$\bm{f}$ are minimized. Nevertheless, it should
be noted that although the notation $D_2$ 
%and
%more generally $D_\alpha$ 
prompts one to think
of derivative operators, the relevant symbolism
of $D_\alpha$ more generally concerns elements
of the nonlinear operator, some of which will,
by necessity, reflect the nonlinearity of the model
(so they should be generally thought of as nonlinear
operators).

In concluding this section, it is worth pointing out that when the dynamical
variables are complex, i.e., $\bu(t)\in \mathbb{C}^N$, they can be decomposed into real and imaginary
parts, i.e., $\bu^{\mathrm{(R)}}$ and $\bu^{\mathrm{(I)}}$, respectively, thus rendering the dynamical variable 
$\bu$ to be a mapping of the form of $\bu:[0,T]\to\mathbb{R}^{2N}$. This way, the PINN $\hat{u}:\R\to\R^{2N}$ is
mapped again through a fully-connected neural network to an output $2N$-dimensional vector now being itself of 
the form of 
$\hat{\bu}(t) = (\hat{u}_1^{\mathrm{(R)}}(t),\hat{u}_2^{\mathrm{(R)}}(t),\dots, \hat{u}_N^{\mathrm{(R)}}(t),%
\hat{u}_1^{\mathrm{(I)}}(t),\hat{u}_2^{\mathrm{(I)}}(t),\dots, \hat{u}_N^{\mathrm{(I)}}(t))\in\R^{2N}$.
%We now move to the next section that discusses our numerical experiments.
Having set up the stage of our computations, we are
now ready to turn to the details of our
numerical experiments.

\section{Numerical Experiments}
\label{sec:num_exp}
% ======================================================================
In all the numerical experiments that we discuss below, we consider the
models summarized in Table~\ref{our_models}. Moreover, we will use $N$, i.e.,
the number of lattice sites, to be in the range $20$ to $31$. %\textcolor{red}
{In our experiments we find that changing the size of the lattice does not seem to change the results in any dramatic way. On the other hand, our experiments suggest that learning is faster when more lattice sites are involved in the dynamics as compared to when the dynamics is local to only a few sites. We made this observation while trying different initial conditions for the $\phi^4$ model where ICs that led to dynamical behavior involving a larger number of lattice nodes led to faster learning and convergence. 
%The initial conditions used
%are summarized in Table~\ref{our_models} and typically
%(but not always or necessarily)
%involve a prototypical solitary wave structure. 
%inspired, e.g.,
%by the continuum limit of the relevant problem
} 

\subsection{Data generation}
\label{sec:data_generation}
% -----------------------------------------------------------------------
At first, we solve the initial-value problems (IVPs) consisting of the models 
of Table~\ref{our_models} and the specific ICs in order to obtain spatio-temporal 
data that will be used for training our PINNs. To do so, we employ temporal
integration. In particular, we use a fourth-order Runge-Kutta (RK4) method 
for the discrete $\phi^4$ and DsG examples, and an implicit backward 
difference scheme~\cite{hairer_wanner_I} for the DNLS and DCGL examples. The ICs 
for our data generation (see, Table~\ref{our_models}) are inspired by the exact solutions 
to the continuous analogues of our models, although
they do not always constitute ones such. For example, and as per the discrete 
$\phi^4$ and DsG cases, we use a traveling kink solution of the respective 
continuum cases.
%(there is an exception in the DsG case). 
% I am not sure what the exception is??
On the other hand, we
use a Gaussian pulse in lieu of a bright soliton for the DNLS %in order to encourage 
%more dynamic behavior in the data generated.
in order to observe an example of dynamics not necessarily
proximal to a solitonic equilibrium.
On the other hand, in the DCGL case, we use a 
form of the so-called Nozaki-Bekki holes \cite{RevModPhys.74.99}. 

As such, we solve the respective IVPs from $t=0$ to $t=10$ with a time step-size
of $dt = 10^{-3}.$ We then extract $50$ samples from the simulated data at equal 
time intervals ($\Delta t = 0.2$), and use these samples to train our neural 
network.

\subsection{Neural Network setup}
% -----------------------------------------------------------------------
We conducted all of our experiments, presented here and otherwise, using the DeepXDE ~\cite{lu2021deepxde} library. 
Our neural networks take as input only time ($t$) and for first-order systems 
output $u_n(t),\,\forall n \in 1, \dots, N,$ while for second-order systems 
output $u_n(t),\,\forall n \in 1, \dots, N,$ and $v_n = \dot u_n(t),\,\forall n \in 1, \dots, N$. 
% For some experiments we employ data augmentation (see Sec. 5 for specifics) 
% and in that case our neural network outputs, depending on the order of the system, 
% either $u_n(t),\,\forall n \in 1, \dots, N,$ or $u_n(t),\,\forall n \in 1, \dots, N,$ 
% and $v_n = \dot u_n(t),\,\forall n \in 1, \dots, N$, for each set of data considered. 
% {\bf PGK: isn't this a bit repetitive? Was it meant to be
% that way?}
Furthermore, as it was already mentioned in Section~\ref{sec:disc_pinns}, in the cases where we have complex data 
(such as in the DNLS and DCGL cases), we split the data into real and imaginary parts 
and learn the two simultaneously, i.e., our neural network outputs: 
$u_n^{\mathrm{R}}(t)$ and $u_{n}^{\mathrm{I}}(t),\,\forall n \in 1, \dots, N$. 

In the residual losses we construct, we only consider the interior nodes (i.e., nodes having both nearest-neighbors). 
This eliminates the need to know the BCs that govern the underlying data. Furthermore, and as per the second-order systems, 
we consider residual losses on both displacements and velocities 
$(u_n,v_n)$, as discussed above.
%$u_n(t),\,\forall n \in 1, \dots, N,$ and $v_n = \dot u_n(t),\,\forall n \in 1, \dots, N$.

Lastly, the neural network architecture used in all the experiments involves fully-connected networks consisting of three hidden 
layers with 40 neurons each, and each layer uses the $\tanh$ activation function.

\subsection{Results and Discussion}
% -----------------------------------------------------------------------

\begin{figure}[!ph]
    \centering
    \begin{overpic}[width=0.4\textwidth]{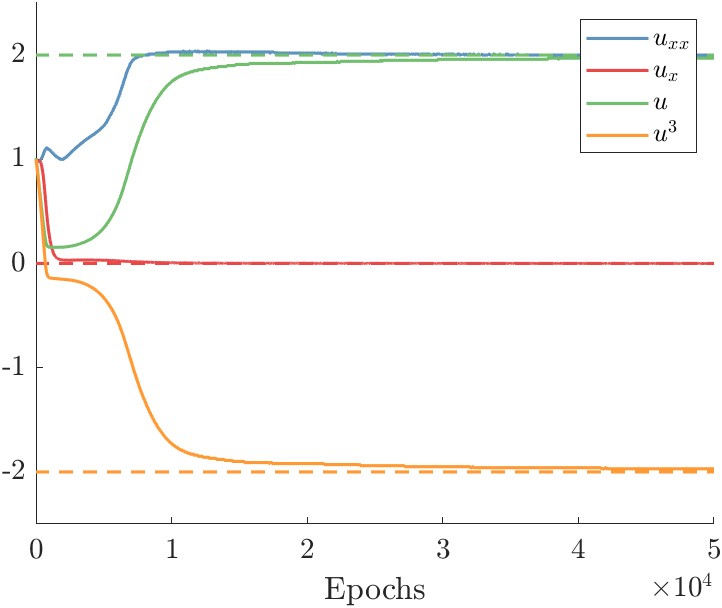}
    \put(50,-4){$(a)$}
     \end{overpic}
     \begin{overpic}[width=0.4\textwidth]{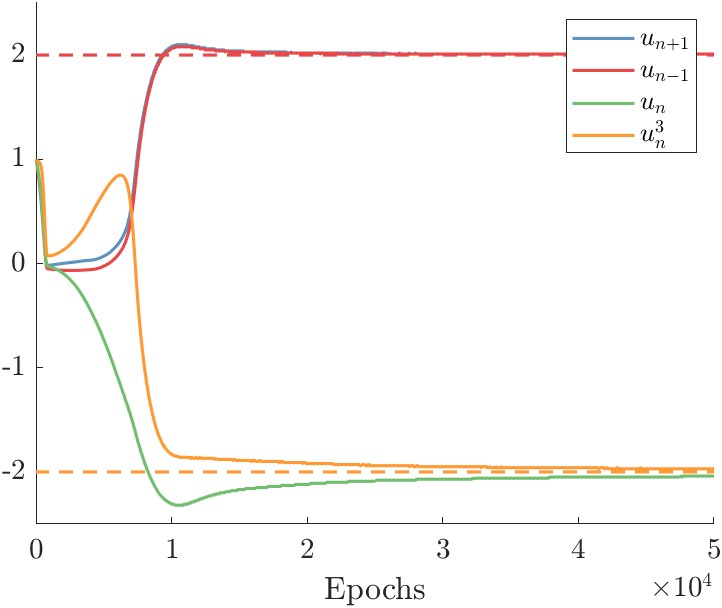}
    \put(50,-4){$(b)$}
     \end{overpic}
     \vskip 0.5cm
     \begin{overpic}[width=0.4\textwidth]{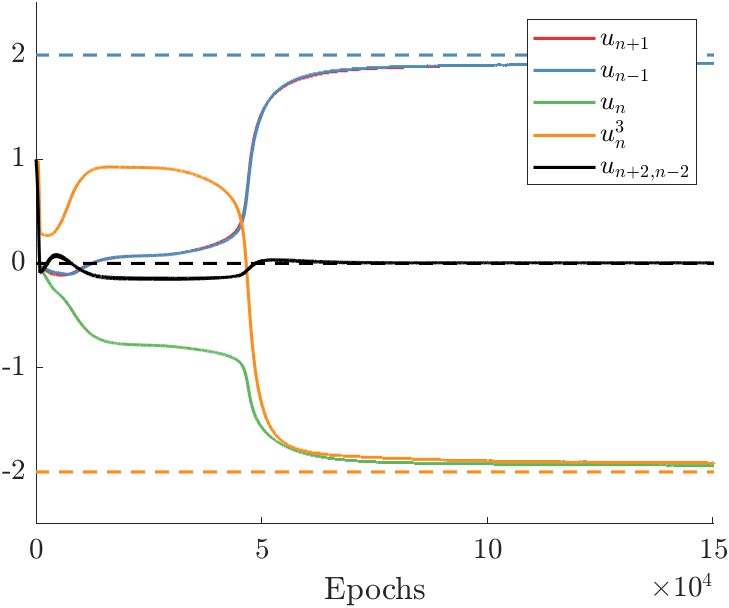}
    \put(50,-4){$(c)$}
     \end{overpic}
     \begin{overpic}[width=0.4\textwidth]{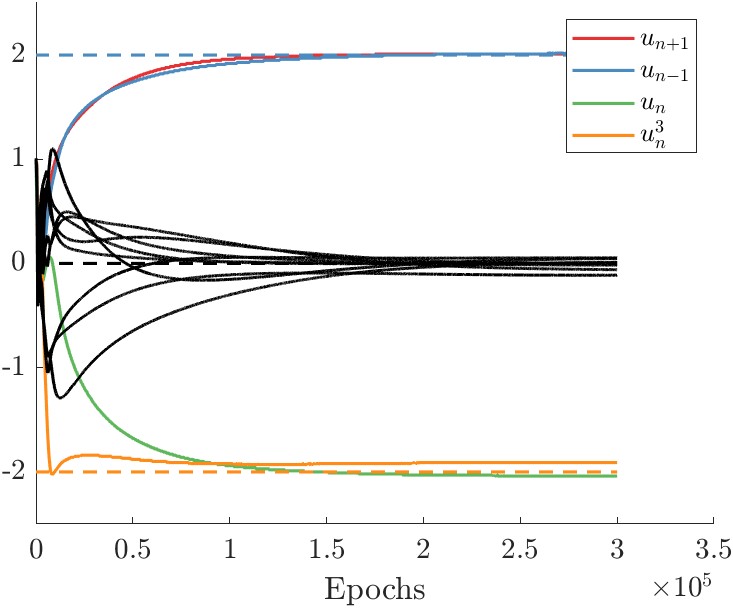}
    \put(50,-4){$(d)$}
     \end{overpic}
    \vspace{0.2cm}
    \caption{Numerical results for the discrete $\phi^{4}$ model [cf.~Eq.~\ref{dphi4}] 
    with $C=2$. In panel (a), the library $\mathrm{Lib}^{(1)}$ of Eq.~\eqref{dphi4_lib1} 
    was considered where the solid blue, red, green and yellow lines correspond to the 
    discrete representation of $u_{xx}$, $u_{x}$, $u$, and $u^{3}$, respectively. The 
    numerical results obtained by using the library $\mathrm{Lib}^{(2)}$ [cf.~Eq.~\eqref{dphi4_lib2} 
    are presented in panel (b) where solid blue, red, green, and yellow depict the 
    $u_{n+1}$, $u_{n-1}$, $u_{n}$, and $u_{n}^{3}$, respectively. The same line-coloring-to-terms 
    correspondence is used in panels (c) and (d) utilizing the libraries of Eqs.~\eqref{dphi4_lib3}
    and~\eqref{dphi4_lib4}, respectively. The solid black lines therein correspond to 
    (c) the terms $u_{n\pm2}$, and (d) to all the other cubic terms. In all the panels,
    the dashed lines serve as reference values for the actual values of the coefficients.}
    \label{fig:exps_dphi4}
\end{figure}

We start our results presentation by the arguably simplest 
among our selected models, namely the (real) discrete $\phi^{4}$ model given by Eq.~\eqref{dphi4}. We consider first
the following library of terms 
\begin{align}
\mathrm{Lib}^{(1)}=\Big\{\left(u_{n+1}+u_{n-1}-2u_n\right),\left(u_{n+1}-u_{n-1}\right)/2,u_n,u_n^3\Big\},
\label{dphi4_lib1}
\end{align}
which contains the discrete Laplacian operator as well as the finite difference representation
of the first derivative (i.e., the second element in Eq.~\eqref{dphi4_lib1} corresponding to a
centered, second-order accurate finite difference operator for $u_{x}$) alongside linear
and cubic terms in $u_{n}$. 
One can argue that this is a library inspired by the
continuum analogue of the model and its respective (potential) 
derivative term inclusions.
Our respective results for $C=2$ are depicted in Fig.~\ref{fig:exps_dphi4}(a)
for this case where the solid red, blue, green and yellow lines correspond to the discrete 
representation of $u_{xx}$, $u_{x}$, $u$, and $u^{3}$, respectively. Indeed, the PINN 
learns the correct coefficients, and most importantly, it learns that there is no $u_{x}$ 
(namely, its discrete version) present in the governing equation for our data.

In the 
experiments that are shown in Figs.~\ref{fig:exps_dphi4}(b)-(d), 
we are taking a more ``inherently discrete'' approach to the relevant
problem. More specifically, motivated by the discreteness
coupling to near-neighbors (rather than to combinations
prompting towards derivatives), we disaggregate
the (discrete) operator $(u_{n+1}+u_{n-1}-2u_n)$ into its constituents. Namely, 
instead of trying to learn the particular form of the inter-component difference, we learn 
the dependence of the governing equation on each of the sets of nearest neighbors. In particular, 
the panels (b), (c), and (d) in the figure consider respectively the following libraries:
\begin{align}
\mathrm{Lib}^{(2)}=\Big\{u_{n+1},u_{n-1},u_{n},u_{n}^{3}\Big\},
\label{dphi4_lib2}
\end{align}
\begin{align}
\mathrm{Lib}^{(3)}=\Big\{\mathrm{Lib}^{(2)},u_{n+2},u_{n-2}\Big\}, 
\label{dphi4_lib3}
\end{align}
and
{\small
\begin{align}
\mathrm{Lib}^{(4)}=\Big\{\mathrm{Lib}^{(2)}, u_{n+1}^2u_n,  %
u_{n-1}^2u_n,u_{n+1}u_{n-1}u_n, u_{n+1}^2u_{n-1}, u_{n-1}^2u_{n+1}, %
u_n^2u_{n+1}, u_n^2u_{n-1}, u_{n+1}^3,  u_{n-1}^3\Big\}.
\label{dphi4_lib4}
\end{align}
}
$\mathrm{Lib}^{(2)}$ is the simplest example containing
the main ingredients of the original model. Hence, one would
like to check whether the methodology can ``disentangle'' the
role of these ingredients from other similar contributions
to both the linear and nonlinear terms. With that in mind,
$\mathrm{Lib}^{(3)}$ is essentially an 
augmentation of $\mathrm{Lib}^{(2)}$ whence the next-nearest neighbors,
i.e., $u_{n\pm2}$ are appended therein. Moreover, $\mathrm{Lib}^{(4)}$ 
contains several possibilities for the cubic nonlinearity of the model.
Indeed, in the latter case all possible combinations 
involving nearest neighbor contributions to a cubic
nonlinearity have been incorporated; see, e.g., 
Ref.~\cite{Dmitriev2009} where also the complex variant
of such terms is discussed in the context of the DNLS model.

In Fig.~\ref{fig:exps_dphi4}(b), we observe that our gradient optimization 
method converges to the correct coefficients, thus constructing the correct
nonvanishing prefactors for each of the relevant term contributions. Next, in the numerical experiments presented in Figs.~\ref{fig:exps_dphi4}(c)-(d),
we consider the libraries of Eqs.~\eqref{dphi4_lib3} and~\eqref{dphi4_lib4}, 
respectively, and try to find the dependence on next-to-next neighbors of each 
node. Here, we expect that solely the relevant ``ingredient''
terms will be selected, while the prefactor of extraneous
contributions will converge to zero.
However, it is important to note, as a limitation of the method,
that for libraries that contained even-ordered terms 
(in particular, quadratic and quartic terms in our experiments),
the model had difficulty learning the correct coefficients, and 
only by using data augmentation,  were we able to get the model to learn the 
correct coefficients. More specifically, we accomplished data augmentation by using the fact that 
if $u$ is a solution to our system, so is $-u$, i.e., leveraging
the relevant invariance of the model under this parity 
transformation of the field. This is, in line, with 
the earlier work of~\cite{zhu_neural_2022}, where we have 
leveraged the symmetries of the model to enhance the network's ability
to solve the inverse problem. 
We thus made the model learn both 
solutions simultaneously, and as expected, the model learned that the governing 
equations do not have any even order terms in them.
In that sense, we have ensured (results not
shown here for brevity) that, using both $u$ and $-u$, even-ordered terms in the library do not alter
the findings presented in Fig.~\ref{fig:exps_dphi4}.

\begin{figure}[!pt]
    \centering
    % \begin{overpic}[width=0.4\textwidth]{DNLS_exp1_c2.png}
    % \put(50,-4){$(a)$}
    %  \end{overpic}
    %  \begin{overpic}[width=0.4\textwidth]{DNLS_exp1_c12.png}
    \begin{overpic}[width=0.4\textwidth]{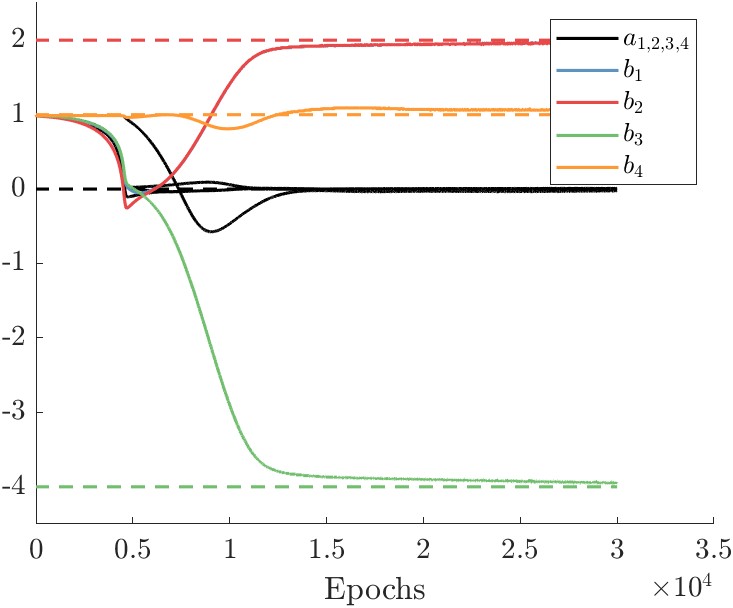}
    \put(50,-4){$(a)$}
     \end{overpic}
     \begin{overpic}[width=0.4\textwidth]{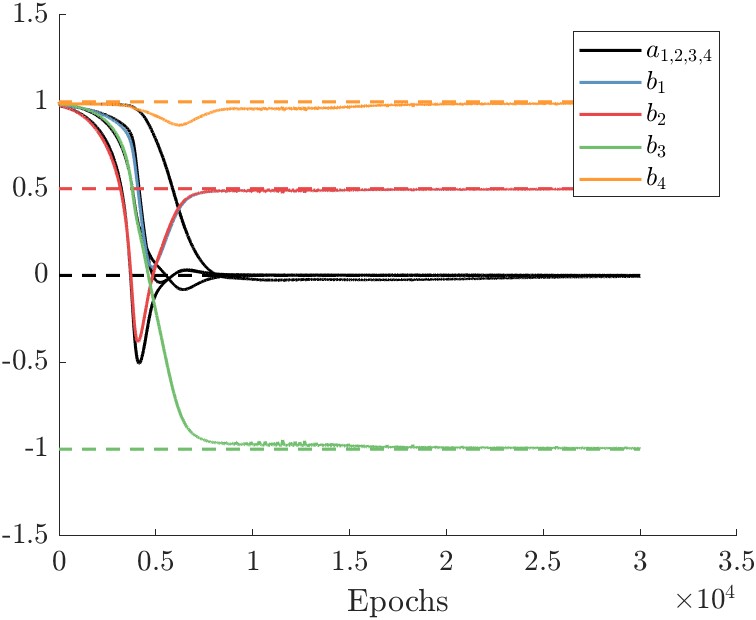}
    \put(50,-4){$(b)$}
     \end{overpic}
    \vspace{0.2cm}
    \caption{Numerical results for the DNLS [cf.~Eq.~\ref{dnls}] with 
    (a) $C=2$ and (b) $C=1/2$ (see, also Eq.~\eqref{dnls_lib_1}). %
    The dashed lines are reference values corresponding to the actual values
    for the coefficients (see text). The solid lines (see the legends of each panel) with colors other than 
    black correspond to the
    imaginary parts of the trained coefficients $b_{1}$ (blue), $b_{2}$ (red), 
    $b_{3}$ (green), and $b_{4}$ (orange). The solid black 
    lines depict the real parts of the trained coefficients ($a_{k}$).}
    \label{fig:exps1_2_dnls}
\end{figure}

%We begin our presentation of numerical results by considering the 
Our next example involved the complex variant of the model,
namely the DNLS [cf.~Eq.~\ref{dnls}], which enables considerable additional richness 
in terms of the available nonlinear terms
(see, e.g., Eq.~(16.11) in~\cite{Dmitriev2009}).
%first, and for various values of $C$. 
It should be noted again that the PINN
models consider real coefficients and thus we split our coefficients herein 
and state variables into real and imaginary parts. As a result, we construct
separate losses for the real and imaginary parts, and set up the PINN to 
learn the respective coefficients simultaneously. Indeed, the panels (a) and 
(b) in Fig.~\ref{fig:exps1_2_dnls} summarize our results herein for $C=2$ 
(a setting more proximal to the continuum limit) and $C=1/2$ (i.e., a rather discrete case),
respectively. For this numerical experiment, we consider a library of the 
form:
\begin{align}
\dot{u}_n = \alpha_1 u_{n+1} + \alpha_2 u_{n-1} + \alpha_3 u_n+ \alpha_4 |u_n|^2u_n,%
\,\,\, \alpha_k = a_k+ib_k\in \mathbb{C}, \,\,\, a_{k},b_{k}\in\R, \,\,\, k=\{1,2,3,4\},
\label{dnls_lib_1}
\end{align}
for both cases (i.e., Figs.~\ref{fig:exps1_2_dnls}(a)-(b)). It can be discerned
from both panels of Fig.~\ref{fig:exps1_2_dnls} that the PINN learns purely imaginary 
coefficients (i.e., $b_{k}$) as expected (see the red, blue, green, and yellow lines 
therein). 
That is, in this case, the scheme detects the effectively
conservative nature of the model, since real coefficients
would be tantamount to gain terms. Consequently, here
the real coefficients $a_{k}$ denoted by solid black lines converge to zero. 
For convenience, in both panels we include the correct values of the coefficients 
for comparison. 

\begin{figure}[!pt]
    \centering
    \begin{overpic}[width=0.4\textwidth]{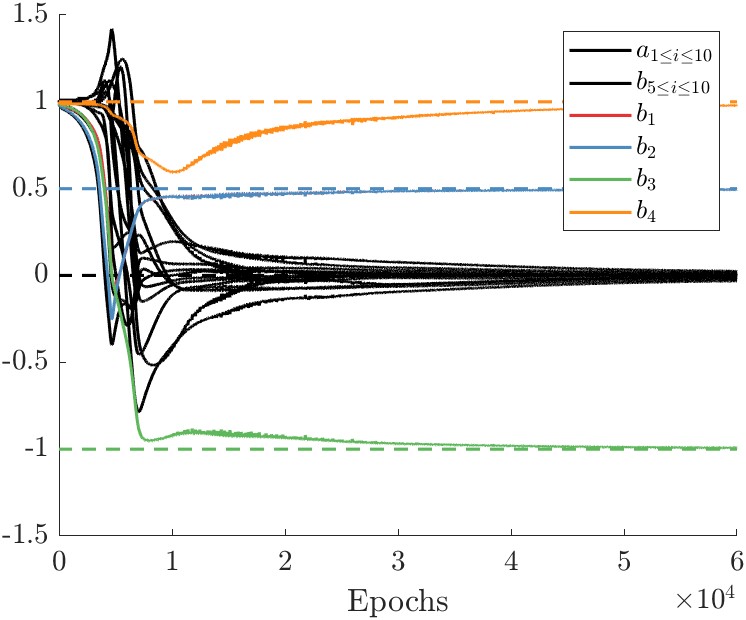}
    \put(50,-4){$(a)$}
     \end{overpic}
     \begin{overpic}[width=0.4\textwidth]{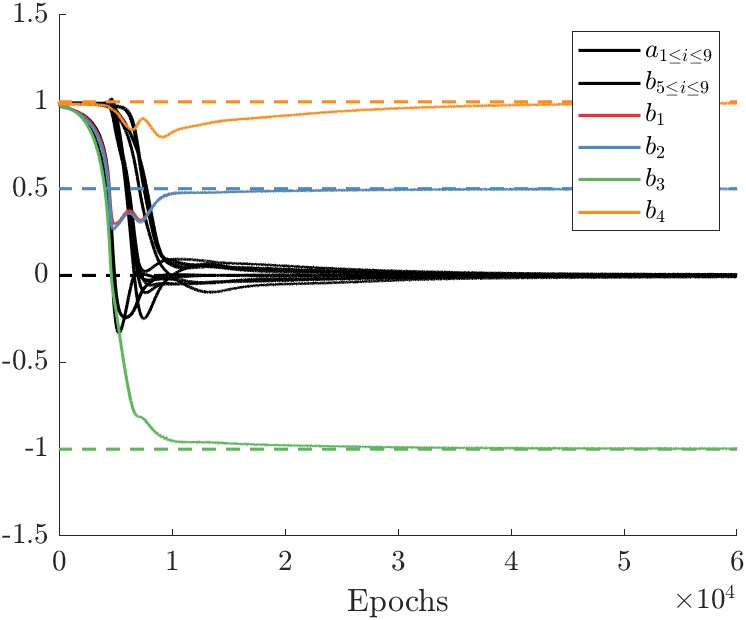}
    \put(50,-4){$(b)$}
     \end{overpic}
    \vspace{0.2cm}
    \caption{Numerical results for the DNLS [cf.~Eq.~\ref{dnls}] both 
    with $C=1/2$, and by using the library of (a) Eq.~\eqref{dnls_lib_3} 
    and (b) Eq.~\eqref{dnls_lib_4}. Similar to Fig.~\ref{fig:exps1_2_dnls},
    the dashed lines are reference values corresponding to the actual values
    for the coefficients. For the line coloring, see the legends of each panel.}
    \label{fig:exps3_4_dnls}
\end{figure}

We further performed numerical experiments on the DNLS with other libraries, motivated by the general cubic nonlinearity
form presented in~\cite{Dmitriev2009} and found that the
PINN is capable of learning the coefficients of the DNLS correctly. Indicatively, we 
demonstrate in Figs.~\ref{fig:exps3_4_dnls}(a)-(b) two cases with $C=1/2$ that, respectively, 
consider the following libraries:
\begin{align}
\dot u_n &=\alpha_1 u_{n+1} + \alpha_2 u_{n-1} + \alpha_3 u_n + \alpha_4 |u_n|^2u_n + \alpha_5 |u_n|^2 + %
           \alpha_6 u_n^2 + \alpha_7 (u_n^{*})^2+\alpha_8 \frac{u_{n+1} + u_{n-1}}{2}u_n  \nonumber \\
         &+\alpha_9 |u_n|^2(u_{n+1}+u_{n-1}) + \alpha_{10} u_n(|u_{n+1}|^2 + |u_{n-1}|^2), %
         \,\,\,\alpha_k = a_k+ib_k\in \mathbb{C},\,\,\, k=\{1,\dots,10\},
         \label{dnls_lib_3}
\end{align}
and 
\begin{align}
\dot u_n &= \alpha_1 u_{n+1} + \alpha_2 u_{n-1} + \alpha_3 u_n + \alpha_4 |u_n|^2u_n + %
          \alpha_5 |u_n|^2 + \alpha_6 |u_{n+1}|^2 + \alpha_7 |u_{n-1}|^2+ \alpha_8 |u_{n+1}|^2u_{n+1}  \nonumber \\
         & + \alpha_9 |u_{n-1}|^2u_{n-1}, \,\,\,\alpha_k = a_k+ib_k\in \mathbb{C},\,\,\, k=\{1,\dots,9\}, 
         \label{dnls_lib_4}
\end{align}
where both $a_{k}$ and $b_{k}$ are real as before. 
Notice that once again here, similarly to the $\phi^4$ case
discussed above, we have included terms that are quadratic
in nature, {and, similarly to the $\phi^4$ case we needed to use data augmentation in order to retrieve the correct coefficients
in the presence of such quadratic terms}.
Despite the generality of the 
above libraries containing themselves different quadratic and cubic nonlinearities, 
the PINN model learned correctly the (purely imaginary) coefficients, as this can be 
discerned from panels (a) and (b) of Fig.~\ref{fig:exps3_4_dnls}. We mention in passing
that we tried other values for the coupling constant $C$ as well as other libraries
(alongside the ones presented so far), and found that in 
all the cases considered 
the PINN discovered the correct coefficients.

\begin{figure}[!pt]
    \centering
    \begin{overpic}[width=0.4\textwidth]{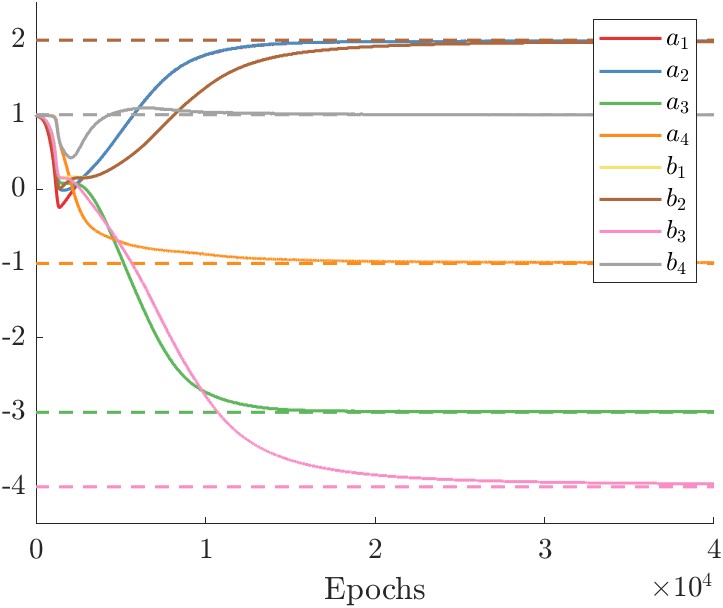}
    \put(50,-4){$(a)$}
     \end{overpic}
     \begin{overpic}[width=0.4\textwidth]{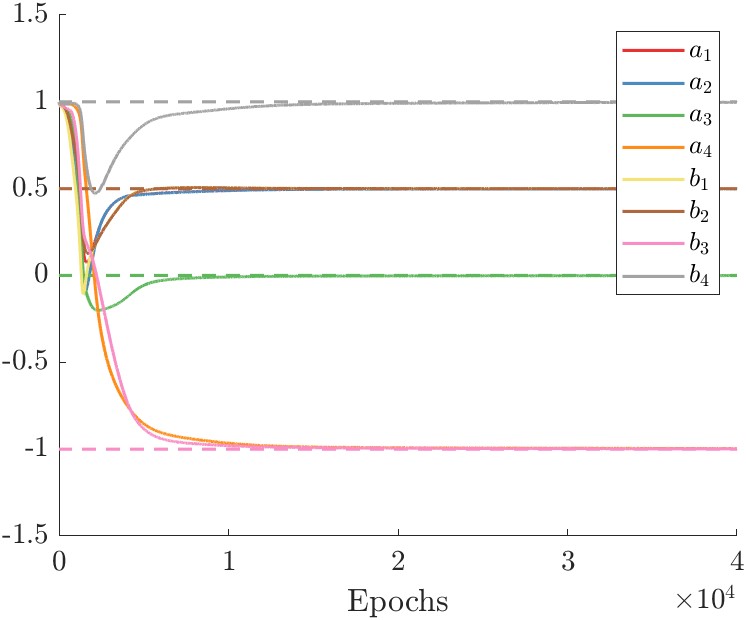}
    \put(50,-4){$(b)$}
     \end{overpic}
    \vspace{0.2cm}
    \caption{Numerical results for the DCGL [cf.~Eq.~\ref{dcgl}] with 
    (a) $C=2$ and (b) $C=1/2$ (see, also Eq.~\eqref{dnls_lib_1}). %
    Same as before, the dashed lines are reference values corresponding 
     to the actual values for the coefficients. 
     Note that $\mathrm{Re}(\alpha_{i})=a_{i}$ and $\mathrm{Im}(\alpha_{i})=b_{i}$
     (see also the legend for the coloring-to-coefficients mapping).}
%      \textcolor{red}{The real parts of the coefficients $\alpha_i$ are 
%      labelled $a_i$ and the imaginary parts are labelled $b_i$}} %\textbf{EGC: Sheikh, you need
    % to clarify the labels here...} }
    \label{fig:exps1_dcgl}
\end{figure}

Having discussed the DNLS, we turn our focus to the DCGL model [cf.~Eq.~\eqref{dcgl}].
Our motivation in doing so was to investigate whether PINNs can learn complex coefficients, when ones such are relevant
for the (general) libraries considered herein. 
At first, we consider the prototypical 
library of Eq.~\eqref{dnls_lib_1}, where we expect to discover
the complex prefactor of both the discrete Laplacian term,
including the equal (complex) coefficients of the $u_{n \pm 1}$
terms, as well as that of the linear term $\propto u_n$.
Finally, the PINN method is able to capture equally accurately
not only the above linear terms, but also the complex
($-(1-i)$)) term of the cubic nonlinearity. 
As is clearly shown in Fig.~\ref{fig:exps1_dcgl}, all the
relevant terms are accurately identified, while the prefactors
of additional, irrelevant terms in the library converge to 
vanishing values beyond a suitably large number of Epochs. {We found this true for a variety of libraries, including ones with even ordered terms and next to next neighbors. We even performed experiments using the libraries of Eq~\ref{dnls_lib_3} Eq~\ref{dnls_lib_4} and found that our models learned the correct coefficients.} 

%\textbf{EGC: I am little bit confused here about the DCGL. 
%Sheikh, please check the original models, the signs etc. because I %tried to reconcile
%the DCGL we write here with what we show in the figures (I am sure, %I am missing something...). 
%I consulted the REU report, and for the DCGL we talk about here, one %has (e.g., for $C=2$): 
%$\alpha_{1,2}=-2+2i$, $\alpha_{3}=4-5i$, $\alpha_{4}=-1-i$. Based on %the experiment 1, I 
%cannot see how these values connect with the figure (plus, more %description is needed about 
%the coloring of the lines--I also tried to find the relevant runs on %github but without success. 
%I would appreciate your help here!). In my opinion, alongside the %above library (and for $C=2$
%and $C=1/2$), we should include the ones of Experiments 3 and 4 %(both for either $C=2$ or $C=1/2$
%as we did for the DNLS) which themselves were considered for the %DNLS (and that would be for 
%homogeneity of the discussion).}

\begin{figure}[!pt]
    \centering
    \begin{overpic}[width=0.4\textwidth]{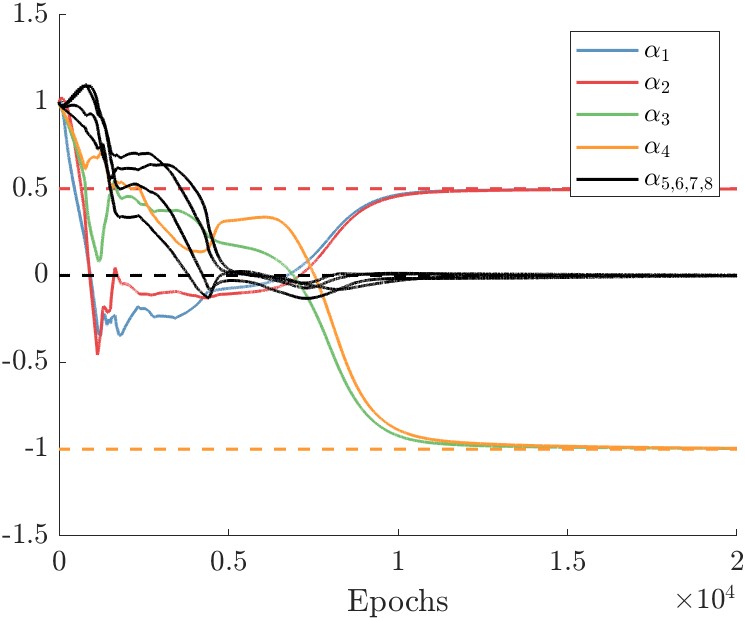}
    \put(50,-4){$(a)$}
     \end{overpic}
     \begin{overpic}[width=0.4\textwidth]{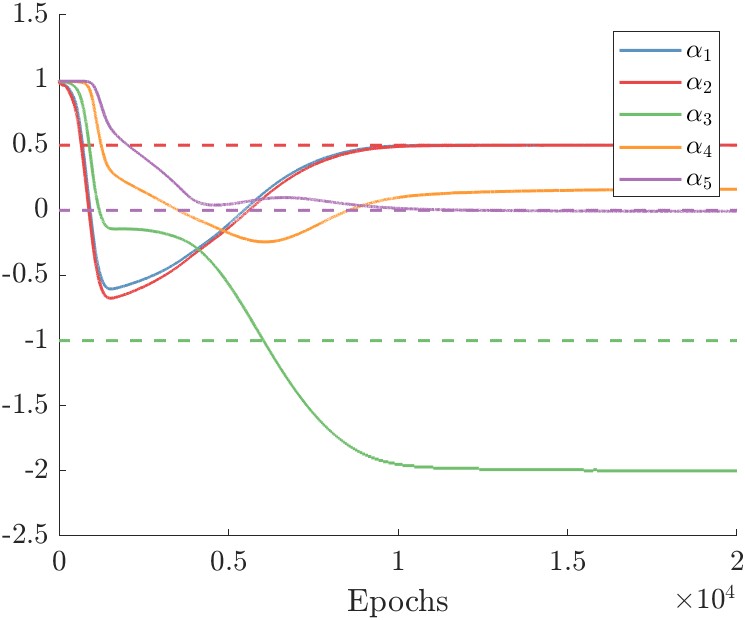}
    \put(50,-4){$(b)$}
     \end{overpic}
     \vskip 0.5cm
     \begin{overpic}[width=0.4\textwidth]{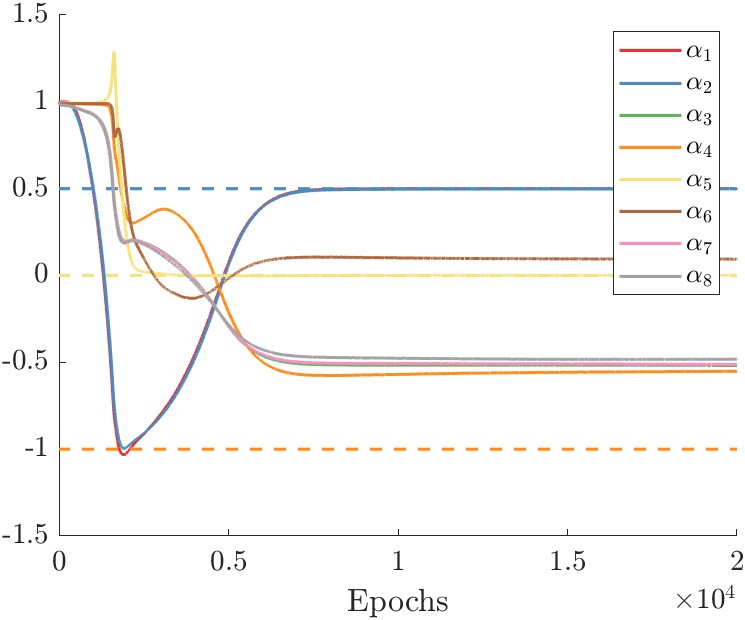}
    \put(50,-4){$(c)$}
    \end{overpic}
    \vspace{0.2cm}
    \caption{Numerical results for the DsG model [cf.~Eq.~\ref{dsG}] with $C=1/2$. 
    In panel (a), the library of Eq.~\eqref{dsG_lib1} was considered whereas panel (b)
    utilized the library of Eq.~\eqref{dsG_lib2}. The numerical results while using the 
    library of Eq.~\eqref{dsG_lib3} are depicted in panel (c). The legends in each of these
    panels offer the line-coloring-to-terms correspondence, and the dashed lines serve as 
    reference values for the actual values of the coefficients.}
%     \caption{Numerical results for the DsG model [cf.~Eq.~\ref{dsG}] with $C=1/2$. 
%     In panel (a), the library of Eq.~\eqref{dsG_lib1} was considered where the 
%     solid red, blue, green and orange lines correspond to the values of $\alpha_{1}$,
%     $\alpha_{2}$, $\alpha_{3}$, and $\alpha_{4}$, respectively. The numerical results 
%     obtained by using the library of Eq.~\eqref{dsG_lib2} are depicted in panel (b),
%     and the line coloring and coefficient values format is the same as in panel (a).
%     Finally, panel (c) considers the library of Eq.~\eqref{dsG_lib3} where the trend 
%     of the coefficients $\alpha_{1}$ (red), $\alpha_{2}$ (blue), $\alpha_{3}$ (green),
%     $\alpha_{4}$ (orange), $\alpha_{5}$ (yellow), $\alpha_{6}$ (brown),
%     $\alpha_{7}$ (pink),
%     $\alpha_{8}$ (grey)
%     respectively, 
%     (over the course of iterations of the optimization solver) is monitored. Similar to 
%     the previous figures, the dashed lines serve as reference values for the actual 
%     values of the coefficients.}
    \label{fig:exps_dsG}
\end{figure}

Finally, 
we choose the discrete sine-Gordon (DsG) model of Eq.~\eqref{dsG} as an intriguing example due to 
the fact that it contains the $\sin\left(u_{n}\right)$ term. The latter can be expanded
in Taylor series, yet it cannot be fully approximated by 
means of a power-law library. It is presumably for this reason
that relevant attempts at the inverse problem of the pendulum~\cite{pmlr-v190-lee22a} or the double 
pendulum~\cite{kaheman} involve libraries containing 
trigonometric terms (rather than purely power-law ones).
This key difference of the present model from the earlier 
ones 
inspired us to use sine series-based libraries, polynomial libraries 
and mixed libraries in order to explore what the PINN model would learn in such a case and what the limitations of each case
example may be. 
For all the numerical experiments that we discuss here, we picked $C=1/2$, and the respective 
results are shown in Fig.~\ref{fig:exps_dsG}. In particular, Fig.~\ref{fig:exps_dsG}(a)
considers the library (once again, effectively, building in the 
$u_n \rightarrow -u_n$ invariance of the model):
{\small
\begin{align}
\ddot{u}_n = \alpha_1 u_{n+1} + \alpha_2 u_{n-1} + \alpha_3 u_n + \alpha_4 \sin{(u_n)} + %
\alpha_5 \sin{(2u_n)} + \alpha_6 \sin{(3u_n)} + \alpha_7 \sin{(4u_n)} + \alpha_8 \sin{(5u_n)}.
\label{dsG_lib1}
\end{align}
}
Here, the PINN model is able to learn the correct sine term (notice that the solid black
lines in the figure correspond to $\alpha_{k}=0,\,\,k=5,6,7,8$, upon convergence). Next, the results presented in Fig.~\ref{fig:exps_dsG}(b) explore the case of a power-law-based
library of functions. Indeed, in this case, we consider
a library that contains three terms of the Taylor series expansion of the sine function
as:
\begin{align}
\ddot{u}_n = \alpha_1 u_{n+1} + \alpha_2 u_{n-1} + \alpha_3 u_n + \alpha_4 u_n^3 + \alpha_5 u_n^5.
\label{dsG_lib2}
\end{align}
It can be discerned from panel (b) of the figure that the model tries to learn a (truncated) 
Taylor series expansion of the sine-term. However, we should mention that the model seems to 
be very sensitive when it comes to polynomial libraries while simultaneously, the number of 
terms in the library seems to have a considerable impact on what the model learns (even when 
all the terms are odd powers). 
Indeed, in this case, for instance, the yellow curve associated with coefficient
$a_4$ converges to a finite value which is clearly distinct
from, e.g., the theoretical Taylor-function prediction of
$1/5!$ We can thus observe the relevant limitation of the
approach in that a polynomial-based library is unable to
fully capture the effects of a sinusoidal nonlinearity.

We conclude our series of experiments by discussing Fig.~\ref{fig:exps_dsG}(c)
which embodies the library:
\begin{align}
\ddot{u}_n = \alpha_1 u_{n+1} + \alpha_2 u_{n-1} + \alpha_3 u_n + \alpha_4 u_n^3 + \alpha_5 u_n^2 %
+ \alpha_6 u_n^5 + \alpha_7 \sin{(u_n)} + \alpha_8 \sin{(2u_n)}, 
\label{dsG_lib3}
\end{align}
i.e., a setting containing both polynomial and sinusoidal terms. Our numerical results suggest that there is some 
sort of a competition between the polynomial and sine terms in trying to describe the nonlinearity 
of the model. It should be noted however, that when trying to learn the coefficients for this model,
the choice of ICs may have a significant impact, especially in
cases of this sort with different competing terms contributing
at the same order. 
We briefly report that  using the exact solution of the 
continuous sine-Gordon seemed to work well for some libraries, and using the exact solution of the 
continuous $\phi^4$ seemed to work well with other libraries. 
{In particular, using the exact solution of the continuous sine-Gordon seems to work better for libraries with $\mathrm{sine}$ terms in them while the exact solution of the continuous $\phi^4$ seems to work better for libraries with polynomial terms corresponding to the Taylor expansion of the $\sin$ nonlinearity.}

\section{Conclusions and Future Challenges}
\label{sec:concl}

In the present work, we have explored the methodology of
Physics-Informed Neural Networks (PINNs) and how PINNs perform when
attempting to solve the inverse problem in the context
of nonlinear dynamical lattices with many degrees of freedom.
We argued herein that, in addition to the relevant problem for
PDEs, such lattice models are of particular interest in their
own right for various physical contexts ranging from optics
to atomic physics to materials science. Hence, a detailed
understanding of the solution of the inverse problem of
coefficient identification is of particular relevance in this
context as well. Indeed, we envision the rather mature
experimental observation and data acquisition techniques
in such settings to be of value in the near future, not
only for machine-learning-based classification 
tasks, as, e.g., has recently been realized
in~\cite{Guo_2021}, but also for data-driven 
modeling efforts.

We started with a simpler real system
case example in the form of the $\phi^4$ model. Here, we were able
to identify the coefficients, although to avoid the possibility
of quadratic terms, a relevant limitation concerned the use
of dynamics both for $u$ and $-u$ to establish the invariance
of the model under such a transformation. Both in the real
case of the $\phi^4$ and in its complex analogue of the 
discrete nonlinear Schr{\"o}dinger lattice,
we considered a wide variety of nonlinear terms. We thus
confirmed that
the additional nonlinearities bear prefactors that eventually
(for sufficiently many epochs) tend to vanishing values,
thereby establishing the models of interest. 
We did not restrict our considerations to purely real
(or purely imaginary) coefficients, but rather extended
them to models with complex ones 
such as the discrete complex Ginzburg-Landau
equation. Finally, we considered cases beyond the setting
of purely power-law nonlinearities, such as the sine-Gordon
lattice. Here, too, we explored some of the limitations
of the PINNs, such as their inability to capture the fully
sinusoidal effects with power-law-based libraries, but also
the potential sensitivity that the concurrent presence of
trigonometric and power-law terms may lead to. 

Naturally, the field of such inverse (and forward) problems
in the realm of nonlinear dynamical lattices is still at a
particularly early stage, and further studies are certainly
warranted. Among the numerous points meriting further exploration,
we note the case example of nonlinearities beyond nearest-neighbors
(and indeed the case of longer-range kernels). Moreover, here
we have restricted considerations to $(1+1)$-dimensional problems,
yet the examination of higher dimensional settings is of 
particular interest in its own right. Beyond these examples,
a progressively deeper understanding of the limitations of the
PINN (or SINDY) type approaches and of how further inclusion
of the model structure (conservation laws, symmetries, 
symplectic nature etc.)
of the underlying system may facilitate convergence are,
in our view, questions of importance for further studies. 
Such efforts are presently in progress and will be reported in
future publications.

\bibliographystyle{unsrt}
\bibliography{sinDdy}

\begin{thebibliography}{10}

\bibitem{Aubry06}
S.~Aubry.
\newblock Discrete breathers: {L}ocalization and transfer of energy in discrete
  {H}amiltonian nonlinear systems.
\newblock {\em Physica D}, 216:1, 2006.

\bibitem{Flach:2008}
S.~Flach and A.V. Gorbach.
\newblock Discrete breathers - {A}dvances in theory and applications.
\newblock {\em Phys. Rep.}, 467(1):1 -- 116, 2008.

\bibitem{FPUreview}
G.~Gallavotti.
\newblock {\em The Fermi--Pasta--Ulam Problem: A Status Report}.
\newblock Springer-Verlag, Berlin, Germany, 2008.

\bibitem{pgk:2011}
P.G. Kevrekidis.
\newblock Non-linear waves in lattices: past, present, future.
\newblock {\em IMA J. Appl. Math.}, 76:389--423, 2011.

\bibitem{moti}
F.~Lederer, G.~I. Stegeman, D.~N. Christodoulides, G.~Assanto, M.~Segev, and
  Y.~Silberberg.
\newblock Discrete solitons in optics.
\newblock {\em Phys. Rep.}, 463:1, 2008.

\bibitem{sievers}
M.~Sato, B.~E. Hubbard, and A.~J. Sievers.
\newblock \textit{Colloquium}: Nonlinear energy localization and its
  manipulation in micromechanical oscillator arrays.
\newblock {\em Rev. Mod. Phys.}, 78:137, 2006.

\bibitem{remoissenet}
M.~Remoissenet.
\newblock {\em Waves Called Solitons}.
\newblock Springer-Verlag, Berlin, 1999.

\bibitem{yuli_book}
Yu. Starosvetsky, K.R. Jayaprakash, M.~A. Hasan, and A.F. Vakakis.
\newblock {\em Dynamics and Acoustics of Ordered Granular Media}.
\newblock World Scientific, Singapore, 2017.

\bibitem{granularBook}
C.~Chong and P.G. Kevrekidis.
\newblock {\em Coherent Structures in Granular Crystals: From Experiment and
  Modelling to Computation and Mathematical Analysis}.
\newblock Springer, New York, 2018.

\bibitem{lars3}
L.~Q. English, M.~Sato, and A.~J. Sievers.
\newblock Modulational instability of nonlinear spin waves in easy-axis
  antiferromagnetic chains. ii. influence of sample shape on intrinsic
  localized modes and dynamic spin defects.
\newblock {\em Phys. Rev. B}, 67:024403, 2003.

\bibitem{ST}
A.~J. Sievers and S.~Takeno.
\newblock Intrinsic localized modes in anharmonic crystals.
\newblock {\em Phys. Rev. Lett.}, 61:970--973, Aug 1988.

\bibitem{page}
J.~B. Page.
\newblock Asymptotic solutions for localized vibrational modes in strongly
  anharmonic periodic systems.
\newblock {\em Phys. Rev. B}, 41:7835--7838, 1990.

\bibitem{alex}
P.~Binder, D.~Abraimov, A.~V. Ustinov, S.~Flach, and Y.~Zolotaryuk.
\newblock Observation of breathers in {Josephson} ladders.
\newblock {\em Phys. Rev. Lett.}, 84:745, 2000.

\bibitem{alex2}
E.~Tr\'{\i}as, J.~J. Mazo, and T.~P. Orlando.
\newblock Discrete breathers in nonlinear lattices: Experimental detection in a
  josephson array.
\newblock {\em Phys. Rev. Lett.}, 84:741, 2000.

\bibitem{Peybi}
M.~Peyrard.
\newblock Nonlinear dynamics and statistical physics of {DNA}.
\newblock {\em Nonlinearity}, 17:R1, 2004.

\bibitem{Yomosa1983}
S.~Yomosa.
\newblock Soliton excitations in deoxyribonucleic acid ({DNA}) double helices.
\newblock {\em Phys. Rev. A}, 27:2120--2125, Apr 1983.

\bibitem{Morsch}
O.~Morsch and M.~Oberthaler.
\newblock Dynamics of {B}ose--{E}instein condensates in optical lattices.
\newblock {\em Rev. Mod. Phys.}, 78:179, 2006.

\bibitem{karniadakis2021physics}
G.E. Karniadakis, I.G. Kevrekidis, L.~Lu, P.~Perdikaris, S.~Wang, and L.~Yang.
\newblock Physics-informed machine learning.
\newblock {\em Nature Reviews Physics}, 3(6):422--440, 2021.

\bibitem{raissi_physics-informed_2019}
M.~Raissi, P.~Perdikaris, and G.~E. Karniadakis.
\newblock Physics-informed neural networks: {A} deep learning framework for
  solving forward and inverse problems involving nonlinear partial differential
  equations.
\newblock {\em Journal of Computational Physics}, 378:686--707, February 2019.

\bibitem{lu2021deepxde}
L.~Lu, X.~Meng, Z.~Mao, and G.E. Karniadakis.
\newblock Deep{XDE}: A deep learning library for solving differential
  equations.
\newblock {\em SIAM Review}, 63(1):208--228, 2021.

\bibitem{brunton_discovering_2016}
S.L. Brunton, J.L. Proctor, and J.N. Kutz.
\newblock Discovering governing equations from data by sparse identification of
  nonlinear dynamical systems.
\newblock {\em Proceedings of the National Academy of Sciences},
  113(15):3932--3937, April 2016.

\bibitem{schaeffer2017learning}
H.~Schaeffer.
\newblock Learning partial differential equations via data discovery and sparse
  optimization.
\newblock {\em Proceedings of the Royal Society A: Mathematical, Physical and
  Engineering Sciences}, 473(2197):20160446, 2017.

\bibitem{feliu2020meta}
J.~Feliu-Faba, Y.~Fan, and L.~Ying.
\newblock Meta-learning pseudo-differential operators with deep neural
  networks.
\newblock {\em Journal of Computational Physics}, 408:109309, 2020.

\bibitem{li2021fourier}
Z.~Li, N.B. Kovachki, K.~Azizzadenesheli, B.~Liu, K.~Bhattacharya, A.~Stuart,
  and A.~Anandkumar.
\newblock Fourier neural operator for parametric partial differential
  equations.
\newblock In {\em International Conference on Learning Representations}, 2021.

\bibitem{sirignano2018dgm}
J.~Sirignano and K.~Spiliopoulos.
\newblock {DGM}: A deep learning algorithm for solving partial differential
  equations.
\newblock {\em Journal of computational physics}, 375:1339--1364, 2018.

\bibitem{weinan2018deep}
E~Weinan and Bing Yu.
\newblock The deep {R}itz method: a deep learning-based numerical algorithm for
  solving variational problems.
\newblock {\em Communications in Mathematics and Statistics}, 6(1):1--12, 2018.

\bibitem{gu2021selectnet}
Y.~Gu, H.~Yang, and C.~Zhou.
\newblock Selectnet: {S}elf-paced learning for high-dimensional partial
  differential equations.
\newblock {\em Journal of Computational Physics}, 441:110444, 2021.

\bibitem{shin2020error}
Y.~Shin, Z.~Zhang, and G.E. Karniadakis.
\newblock Error estimates of residual minimization using neural networks for
  linear {PDE}s.
\newblock {\em arXiv preprint arXiv:2010.08019}, 2020.

\bibitem{luo2020two}
T.~Luo and H.~Yang.
\newblock Two-layer neural networks for partial differential equations:
  {O}ptimization and generalization theory.
\newblock {\em arXiv preprint arXiv:2006.15733}, 2020.

\bibitem{zhu_neural_2022}
W.~Zhu, W.~Khademi, E.G. Charalampidis, and P.G. Kevrekidis.
\newblock Neural {Networks} {Enforcing} {Physical} {Symmetries} in {Nonlinear}
  {Dynamical} {Lattices}: {The} {Case} {Example} of the {Ablowitz}-{Ladik}
  {Model}.
\newblock {\em Physica D: Nonlinear Phenomena}, 434:133264, June 2022.

\bibitem{george_again}
P.~Jin, Z.~Zhang, A.~Zhu, Y.~Tang, and G.E. Karniadakis.
\newblock {S}ymp{N}ets: {I}ntrinsic structure-preserving symplectic networks
  for identifying {H}amiltonian systems.
\newblock {\em Neural Networks}, 132:166--179, 2020.

\bibitem{p4book}
J.~Cuevas-Maraver and P.~G.~Kevrekidis (eds.).
\newblock {\em A dynamical perspective on the $\phi^4$ model}.
\newblock Nonlinear Systems and Complexity. Springer International Publishing,
  1st edition, 2019.

\bibitem{kevrekid_dnls_book}
P.G Kevrekidis.
\newblock {\em {D}iscrete {N}onlinear {S}chr\"odinger {E}quation:
  {M}athematical {A}nalysis, {N}umerical {C}omputations and {P}hysical
  {P}erspectives}.
\newblock Springer-Verlag (Heidelberg, 2009).

\bibitem{braun1998}
O.M. Braun and Y.S. Kivshar.
\newblock Nonlinear dynamics of the {{Frenkel}}-{{Kontorova}} model.
\newblock {\em Physics Reports}, 306(1-2):1--108, December 1998.

\bibitem{SGbook}
J.~Cuevas-Maraver, P.~G. Kevrekidis, and F.~L.~Williams (eds.).
\newblock {\em The {sine-Gordon} Model and its Applications: From Pendula and
  {Josephson} Junctions to Gravity and High-Energy Physics}.
\newblock Nonlinear Systems and Complexity 10. Springer International
  Publishing, 1st edition, 2014.

\bibitem{PhysRevE.67.026606}
N.K. Efremidis and D.N. Christodoulides.
\newblock {D}iscrete {G}inzburg-{L}andau solitons.
\newblock {\em Phys. Rev. E}, 67:026606, Feb 2003.

\bibitem{GL2}
N.K. Efremidis, D.N. Christodoulides, and K.~Hizanidis.
\newblock Two-dimensional discrete {G}inzburg-{L}andau solitons.
\newblock {\em Phys. Rev. A}, 76:043839, Oct 2007.

\bibitem{RevModPhys.74.99}
I.S. Aranson and L.~Kramer.
\newblock The world of the complex {G}inzburg-{L}andau equation.
\newblock {\em Rev. Mod. Phys.}, 74:99--143, Feb 2002.

\bibitem{Salerno2022}
M.~Salerno and F.Kh. Abdullaev.
\newblock {\em Discrete Solitons of the Ginzburg-Landau Equation}, pages
  303--317.
\newblock Springer International Publishing, Cham, 2022.

\bibitem{PhysRevE.76.026601}
I.~Roy, S.V. Dmitriev, P.G. Kevrekidis, and A.~Saxena.
\newblock Comparative study of different discretizations of the
  ${\ensuremath{\phi}}^{4}$ model.
\newblock {\em Phys. Rev. E}, 76:026601, Aug 2007.

\bibitem{PhysRevE.72.035602}
I.~V. Barashenkov, O.~F. Oxtoby, and Dmitry~E. Pelinovsky.
\newblock Translationally invariant discrete kinks from one-dimensional maps.
\newblock {\em Phys. Rev. E}, 72:035602, Sep 2005.

\bibitem{pmlr-v190-lee22a}
K.~Lee, N.~Trask, and P.~Stinis.
\newblock Structure-preserving sparse identification of nonlinear dynamics for
  data-driven modeling.
\newblock In B.~Dong, Q.~Li, L.~Wang, and Z.-Q.~J. Xu, editors, {\em
  Proceedings of Mathematical and Scientific Machine Learning}, volume 190 of
  {\em Proceedings of Machine Learning Research}, pages 65--80. PMLR, 15--17
  Aug 2022.

\bibitem{doi:10.1126/science.abf6873}
S.~Xia, D.~Kaltsas, D.~Song, I.~Komis, J.~Xu, A.~Szameit, H.~Buljan, K.G.
  Makris, and Z.~Chen.
\newblock Nonlinear tuning of pt symmetry and non-hermitian topological states.
\newblock {\em Science}, 372(6537):72--76, 2021.

\bibitem{doi:10.1126/sciadv.aat6539}
A.~Müllers, B.~Santra, C.~Baals, J.~Jiang, J.~Benary, R.~Labouvie, D.A.
  Zezyulin, V.V. Konotop, and H.~Ott.
\newblock Coherent perfect absorption of nonlinear matter waves.
\newblock {\em Science Advances}, 4(8):eaat6539, 2018.

\bibitem{baydin2018automatic}
A.G. Baydin, B.A. Pearlmutter, A.A Radul, and J.M. Siskind.
\newblock Automatic differentiation in machine learning: a survey.
\newblock {\em Journal of machine learning research}, 18, 2018.

\bibitem{DBLP:journals/corr/KingmaB14}
D.P. Kingma and J.~Ba.
\newblock Adam: {A} {M}ethod for {S}tochastic {O}ptimization.
\newblock In {\em ICLR (Poster)}, 2015.

\bibitem{liu1989limited}
D.C. Liu and J.~Nocedal.
\newblock On the limited memory {BFGS} method for large scale optimization.
\newblock {\em Mathematical programming}, 45(1):503--528, 1989.

\bibitem{hairer_wanner_I}
E.~Hairer, S.P. N\o{}rsett, and G.~Wanner.
\newblock {\em {S}olving ordinary differential equations I. {N}onstiff
  {P}roblems}.
\newblock Springer-Verlag (Heidelberg, 2008).

\bibitem{Dmitriev2009}
S.V. Dmitriev and A.~Khare.
\newblock {\em Exceptional Discretizations of the NLS: Exact Solutions and
  Conservation Laws}, pages 293--310.
\newblock Springer Berlin Heidelberg, Berlin, Heidelberg, 2009.

\bibitem{kaheman}
K.~Kaheman, J.N. Kutz, and S.L. Brunton.
\newblock Sindy-pi: a robust algorithm for parallel implicit sparse
  identification of nonlinear dynamics.
\newblock {\em Proc. R. Soc. A}, 476:20200279, 2020.

\bibitem{Guo_2021}
Shangjie Guo, Amilson~R Fritsch, Craig Greenberg, I~B Spielman, and Justyna~P
  Zwolak.
\newblock Machine-learning enhanced dark soliton detection in bose–einstein
  condensates.
\newblock {\em Machine Learning: Science and Technology}, 2(3):035020, jun
  2021.

\end{thebibliography}

\end{document}